# Unmasking the giant: A comprehensive evaluation of ChatGPT's proficiency in coding algorithms and data structures

SAYED ERFAN AREFIN, Texas Tech University, USA
TASNIA ASHRAFI HEYA, Texas Tech University, USA
HASAN AL-QUDAH, Texas Tech University, USA
YNES INEZA, Texas Tech University, USA
ABDUL SERWADDA, Texas Tech University, USA

The transformative influence of Large Language Models (LLMs) is profoundly reshaping the Artificial Intelligence (AI) technology domain. Notably, ChatGPT distinguishes itself within these models, demonstrating remarkable performance in multi-turn conversations and exhibiting code proficiency across an array of languages. In this paper, we carry out a comprehensive evaluation of ChatGPT's coding capabilities based on what is to date the largest catalog of coding challenges. Our focus is on the python programming language and problems centered on data structures and algorithms, two topics at the very foundations of Computer Science. We evaluate ChatGPT for its ability to generate correct solutions to the problems fed to it, its code quality, and nature of run-time errors thrown by its code. Where ChatGPT code successfully executes, but fails to solve the problem at hand, we look into patterns in the test cases passed in order to gain some insights into how wrong ChatGPT code is in these kinds of situations. To infer whether ChatGPT might have directly memorized some of the data that was used to train it, we methodically design an experiment to investigate this phenomena. Making comparisons with human performance whenever feasible, we investigate all the above questions from the context of both its underlying learning models (GPT-3.5 and GPT-4), on a vast array sub-topics within the main topics, and on problems having varying degrees of difficulty.

## 1 INTRODUCTION

Artificial Intelligence has made remarkable strides, showing extraordinary potential in automating a wide range of sectors. A significant contributor to this progress has been the advancement in Large Language Models (LLMs), which have essentially transformed the landscape of AI technology. These models are trained to comprehend and generate text that closely mimics human communication, responding intelligently based on the input they receive. Among these models, ChatGPT has emerged as a standout, particularly due to its capacity for sustained, multi-turn conversations. This enables it to maintain coherence and context over extended dialogues.

In addition to drafting and refining content, exploring new topics, and brainstorming creative ideas, ChatGPT has made a considerable impact in the realm of software development. Its capabilities extend beyond generating code in multiple languages. ChatGPT can provide explanatory commentary for code snippets, review existing code, and identify potential issues or bugs. These functionalities have led to a wealth of applications. For instance, it can serve as an instructive tool for novice programmers, assist in code reviews, and expedite the prototyping process, to name just a few uses. As the emergence of new use-cases continues at an impressive rate, there's a growing necessity for in-depth analysis of ChatGPT's coding capabilities. By understanding its strengths and limitations, users can be given essential guidance which can in turn enable them to make

Authors' addresses: Sayed Erfan Arefin, saarefin@ttu.edu, Texas Tech University, Lubbock, Texas, USA, saarefin@ttu.edu; Tasnia Ashrafi Heya, tasnia.heya@ttu.edu, Texas Tech University, Lubbock, Texas, USA; Hasan Al-Qudah, halqudah@ttu.edu, Texas Tech University, Lubbock, Texas, USA; Ynes Ineza, Texas Tech University, Lubbock, Texas, USA, yineza@ttu.edu; Abdul Serwadda, abdul.serwadda@ttu.edu, Texas Tech University, Lubbock, Texas, USA.

the most of this groundbreaking family of technologies. Simultaneously, this understanding can highlight areas of improvement to the developers, ensuring the continual evolution of these tools.

In this paper, we carry out an evaluation of ChatGPT's coding capabilities based on what is to date the largest catalog of coding challenges. Our focus is on the python programming language, the most popular language in a growing number of fields, including artificial intelligence, machine learning, data analytics, automation, and, scientific computing [17]. All challenges solved in our research are centered on data structures and algorithms, two topics at the very foundations of Computer Science. Most real-world programming tasks involve some aspect of algorithms and data structures [10]. Whether it's organizing data for efficient access (using appropriate data structures), or creating workflows to process data (using effective algorithms), these concepts are fundamental to daily programming. It is for these same reasons that these topics are at the core of evaluations of human candidate abilities during coding job interviews [4]. By virtue of the number and diversity of coding challenges posed to ChatGPT, variety of scenarios studied and attributes evaluated, *this paper is to our knowledge the most comprehensive evaluation of ChatGPT's coding proficiency* in the algorithms and data structures space to date.

We evaluate ChatGPT not only for its ability to generate correct solutions to the problems fed to it, but also for its code quality, and nature of run-time errors thrown by its code. Where ChatGPT code successfully executes, but fails to solve the problem at hand, we compile statistics, based on test cases evaluated, that provide some insights into how wrong ChatGPT code is in these kinds of situations. To gain some insights into whether ChatGPT might have directly memorized some of the data that was used to train it, we methodically design an experiment to investigate this phenomena. We investigate all these questions from the context of both its publicly available underlying models (GPT-3.5[1] and GPT-4), a vast array sub-topics within the main topics, and varying degrees of difficulty of the problems.

The paper makes the following four primary contributions:

**Evaluating correctness of ChatGPT coding solutions across a diverse spectrum of algorithms and data structures problems**: Utilizing a comprehensive set of 2,792 coding prompts, we evaluate the accuracy of coding solutions produced by ChatGPT, spanning a diverse array of algorithms and data structures topics. Our analysis explores five key subtopics within algorithms: dynamic programming, greedy algorithms, depth-first search, divide and conquer, and topological sort. Concurrently, we investigate five fundamental areas within data structures, namely, priority queues, arrays, hash tables, stacks, and binary search trees. In addition, we conduct a detailed examination of string manipulation. This is a topic that intersects both data structures and algorithms, yet it boasts its own distinctive nuances, particularly concerning operations, and various string-specific data structures and algorithms. Central to our experiments is LeetCode [6], a platform renowned for its diverse collection of coding challenges. LeetCode's built-in compiler enables the evaluation of submitted solutions. The use of this platform not only facilitates an in-depth assessment of the correctness of a coding submission but also allows us to compare each solution against a historical catalog of solutions submitted by millions of developers over the years. In this way, we are able to address the increasingly prevalent question: How does AI performance measure up against human performance?

**Evaluation of ChatGPT's code quality**: Beyond the correctness of a coding solution, another key measure of coding proficiency is the code quality, a characterization of how well code is written in relation to well-established good practices of programming. For this evaluation, we use PyLint [13], a widely-used tool in the Python programming language for checking a module for coding standards, and certain types of code smells. We use this to meticulously evaluate ChatGPT's code for details of warnings, adherence to conventions, refactoring hints, and errors. We carry out this analysis regardless of whether ChatGPT provides a wrong or correct solution to the problem at hand, since the notion of code quality continues to be insightful either way. Whether it is a beginner programmer using ChatGPT during their learning activities, a more experienced developer importing segments of ChatGPT-generated code into a project, or a researcher looking into areas to improve LLMs such as ChatGPT, these code quality assessments should offer some useful insights.

---

[1] In parts of the paper, we also refer to this by the base model number, GPT-3

**Examining ChatGPT for potential memorization of training data**: One of the fears surrounding LLMs is that they might memorize (potentially private) data that shows up in the training set. We design an experiment to provide some idea of how ChatGPT might be affected by this problem. In particular, we subject ChatGPT to a series of incomplete coding challenges in which critical information was left out. These questions are drawn from part of the dataset that was used to train ChatGPT, albeit with the aforementioned pieces of information left out. We find ChatGPT to generate correct code for some of these challenges, suggesting that it might have memorized these problems and their solutions. While this experiment would not confirm 100% whether memorization has occurred, it can still provide some perspective to the end-user who is trying to make the decision on whether to allow their data to be used in the training set. The contribution is also of value to the AI community who are working to develop and tune LLMs such as ChatGPT.

**Assessing the level of "wrongness" of wrong ChatGPT solutions**: When ChatGPT generates a wrong solution to a problem, it is still insightful to gauge the extent to which this solution is wrong. For example, a marginally wrong solution might be fixable through minor tweaks. To study this phenomena, we evaluated the test cases passed by ChatGPT-generated programs which executed successfully yet failed to solve the problem at hand. Findings on this question should be informative to not only experienced programmers seeking to import ChatGPT code into their projects but also beginners using ChatGPT as a programming tutor.

**Paper Organization:** The rest of the paper is organized as follows. We discuss related research in Section 2 and our data collection experiments in Section 3. We then present our findings in Section 4, and conclusions in Section 5.

## 2 RELATED RESEARCH

We categorize past research into two streams, namely: (1) research that evaluated ChatGPT's coding capabilities, and, (2) research that evaluated ChatGPT's performance on tasks other than coding. Below, we discuss these research streams and how our work differs from them.

### 2.1 Research on ChatGPT's coding capabilities

Noever et al. [11], conducted a study to assess the ability of ChatGPT to conduct CRUD (Create Read Update Delete) operations on data science datasets. This evaluation utilized four well-known datasets: the Iris [1], Titanic [7], Boston Housing[15], and Faker [14] datasets. The study showed that ChatGPT successfully emulated a Python interpreter, autonomously generating code and delivering the expected output. These results suggest ChatGPT possesses the necessary competencies to manage structured datasets and execute CRUD operations effectively. Another study by Biswas et al. [3] underlined ChatGPT's potential as an intelligent assistant to programmers. In this study, ChatGPT demonstrated the ability to provide support in programming tasks such as code completion, bug fixing, and code refactoring among others. The study also showed that ChatGPT could synthesize code snippets based on predefined specifications, consequently minimizing the manual effort typically associated with code development from scratch.

In a study employing a methodology similar to ours [5], ChatGPT was tasked to write Python functions. Initially, the HumanEval dataset was leveraged to gauge ChatGPT 's programming ability. To prevent any potential effects of memorization from tainting the results, additional evaluation was done using a set of 100 LeetCode problems which, at the time of ChatGPT's training completion, were not available on the Internet. Performance of GPT-4 was bench-marked against other models as well as against human performance metrics derived from LeetCode contest results. The results demonstrated that GPT-4 performs better in comparison to the other LLM models such as text-davinci-003 (the base model of ChatGPT), Codex (code-davinci-002), and CODEGEN-16B.

Table 1. Comparing our research with the sub-set of related works that are most similar to our work

| Research focus | | Publications | | | | |
| --- | --- | --- | --- | --- | --- | --- |
| | | Noever et al.[11] | Biswas et al.[3] | Tian et al.[16] | Bubeck et al.[5] | This work |
| Experiment details | Number of coding prompts | Not specified | Not specified | 240 | 264 | 2,792 |
| | Separate train and test problems | | | ✓ | | ✓ |
| Coding problems solved | Dynamic algorithms | | | ✓ | ✓ | ✓ |
| | Greedy algorithms | | | | | ✓ |
| | Depth first search | | | | | ✓ |
| | Divide and conquer | | | | | ✓ |
| | Topological sort | | | | | ✓ |
| | Priority queue | | | | | ✓ |
| | Arrays | | | ✓ | | ✓ |
| | Hash tables | | | ✓ | | ✓ |
| | Stacks | | | | | ✓ |
| | Binary search trees | | | | | ✓ |
| | Strings | | | ✓ | | ✓ |
| Attributes of ChatGPT solutions evaluated | Correctness of solutions | ✓ | | ✓ | ✓ | ✓ |
| | Runtime errors | | | | | ✓ |
| | Memorization of trainset | | | ✓ | | ✓ |
| | Analysis of wrong solutions | | | | | ✓ |
| Code quality evaluation | Error classification | | | | | ✓ |
| | Warnings | | | | | ✓ |
| | Conventions | | | | | ✓ |
| | Refactoring | | | | | ✓ |
| Model evaluated | GPT-3 | ✓ | ✓ | ✓ | | ✓ |
| | GPT-4 | | | | ✓ | ✓ |

In a related investigation [16], researchers undertook a comprehensive three-pronged assessment of ChatGPT's capabilities. First, they probed its ability to translate natural language into code, using two LeetCode datasets containing questions of varied complexity as prompts. The second phase of the evaluation aimed to ascertain ChatGPT's utility as a code rectification tool, tasking it with the correction of a broad spectrum of incorrect codes sourced from a programming assignments benchmark. The final aspect of the study scrutinized ChatGPT's capacity to reason on both correct and incorrect code samples drawn from a student assignment benchmark, thereby testing its code comprehension and interpretive abilities. The authors of this paper also compared the performance of the coding solutions generated by ChatGPT against other models like Codex (Dex), and CodeGen (Gen).

While both works in [5] and [16] demonstrated the commendable reasoning and coding capabilities of ChatGPT, they also highlighted the model's shortcomings. It was observed that ChatGPT, while generally proficient, can occasionally falter by generating erroneous codes, or those that are semantically incorrect or syntactically invalid. Moreover, its effectiveness can be compromised when dealing with more intricate programming tasks, suggesting a potential struggle with comprehensive understanding of complex instructions or domain-specific problems.

In [17], a crowdsourcing data-driven framework to investigate the code generation performance of ChatGPT was presented. The paper employed a hybrid keyword expansion method to filter relevant social media posts about ChatGPT's code generation on platforms like Twitter and Reddit, resulting into the collection of 316,000 tweets and 3,200 Reddit posts spanning from December 1, 2022, to January 31, 2023. Analysis of these tweets provided answers to a number of questions, including: the most popular programming languages in ChatGPT usage, programming scenarios, tasks, and purposes that people use ChatGPT for, temporal distribution of the discussion on ChatGPT code generation, stakeholders' perception of ChatGPT code generation, and nature of errors in the Python code snippets generated by ChatGPT.

These papers align with our fundamental aim to delineate the coding capabilities of ChatGPT. However, our research diverges from these studies in several significant respects. These differences include the scale of our experiments, the types of coding tasks employed, the specific coding modalities examined, and the variety of language models evaluated. We delve into these distinctions next. For the four papers most similar to our work, a summary of some of these variances can also be viewed in Table 1.

(1) *Size of the experiment*: Our study is based on a total of 2,792 coding challenges distributed across the various experimental settings used in our study. To the best of our knowledge, this constitutes the most extensive study to date that investigates ChatGPT's coding capabilities. By comparison, the largest study (see [5]) among the aforementioned related works utilized a total of 264 coding prompts. In experimental research of this nature, the magnitude of the experiment is crucial in bolstering the statistical significance of the results. Through exposure to a wide array of prompts, one can better stimulate and scrutinize the diverse potential strengths and limitations of the technology in question.

(2) *Variety of coding tasks used in experiment*: Our work notably expands upon the studies by Bubeck et al. and Tian et al., which are closest to ours in terms of focusing on programming tasks concerning data structures and algorithms. Tian et al. concentrated on two specific subtopics, namely arrays and hash tables within data structures, and sorting within algorithms. Bubeck et al., on the other hand, devoted their study to dynamic programming, a singular subtopic of algorithms. Our research, however, extends far beyond these areas. We delve into five subtopics under algorithms: dynamic programming, greedy algorithms, depth first search, divide and conquer, and topological sort. Simultaneously, we explore five areas within data structures: priority queues, arrays, hash tables, stacks, and binary search trees. By broadening the scope of our study, we aim to offer the community significant insights into ChatGPT's coding proficiency across a wide array of fundamental Computer Science problems not previously explored. This comprehensive approach should further illuminate the platform's capabilities.

(3) *Coding quality evaluation*: Most of the above cited works have primarily concentrated on the correctness of the final coding solutions produced by the language models. Our approach, however, encompasses more than this criterion. We evaluate both GPT-3 and GPT-4 against a comprehensive range of quality metrics, including but not limited to, coding conventions, intricacies of coding errors, warnings, and refactoring. Such analysis is of paramount importance for numerous reasons. Language models like ChatGPT are increasingly becoming the go-to tools for budding programmers, while even seasoned developers are incorporating extensive blocks of GPT-produced code into their projects. As these tools continue to evolve, our analysis will be instrumental in guiding enhancements that will render them even more beneficial to the user community. The work in [17] includes some statistics on coding errors (as assessed by Flake8), but has a fundamentally different thrust from us given its focus on leveraging the chatter generated on social media to answer questions about ChatGPT's code generation and general usage.

(4) *Language models under evaluation*: Our work is focused on GPT-4 and its predecessor GPT-3 (we use model 3.5 particular), which are the state-of-the art models on the consumer market via ChatGPT. We analyze each of them individually, and also compare them against each other based on exactly the same coding challenges. The works in [11], [3], and [16], having been published before GPT-4 came out, focused on evaluating GPT-3, and how it compares to older models such as Codex and CodeGen. The closest to us is the work in [5] since it also performs coding performance evaluations on GPT-4. However, this work is put apart from our work by several earlier described features, namely, the size of the experiment, variety of computing problems solved and the fact that they do not undertake coding quality evaluations.

(5) *Training set memorization and assessment of wrong solutions*: Finally, the evaluation of ChatGPT's memorization behavior and the assessment of the extent of "wrongness" of ChatGPT's wrong solutions (recall Section 1) are novelties in our work that have not been studied by any of the previous works on ChatGPT.

## 2.2   Research on ChatGPT's performance on tasks other than coding

Much more distant from our research is the body of work which has evaluated ChatGPT on tasks other than coding.

For example, in [9], ChatGPT was trained on medical texts and assessed for its ability to answer United States Medical Licensing Examination (USMLE) questions. It achieved an average accuracy of 61.2%, lower than human test-takers. It was found to perform well on recall-based questions, but poorly on reasoning-based questions. The works in [5, 8, 12] evaluated ChatGPT in the domain of mathematical understanding. In [5], it was found that the model showed proficiency in creative reasoning but lacked critical reasoning skills. It demonstrated technical proficiency in algorithms but made frequent mistakes in calculations and notation. The authors concluded that while further training might help alleviate some issues, fundamental limitations remained.

Frieder et al. [12] developed the GHOSTS dataset containing graduate-level math test questions. They evaluated ChatGPT's math abilities on this dataset. In the Grad Text dataset, ChatGPT performed well, but in the Olympiad-Problem-Solving and Holes-in-Proofs datasets, it scored poorly. In the MATH dataset, ChatGPT provided correct solutions in only 26% of cases. These findings indicate that ChatGPT's math abilities are inferior to those of math graduate students. While ChatGPT can comprehend math problems, it struggles to generate correct solutions. The researchers in [8] compared the effectiveness of hints generated by ChatGPT and human tutors in algebra topics: Elementary and Intermediate algebra. However, the learning gains from human-authored hints were significantly higher than those from ChatGPT hints in both topic areas.

The works in [2, 18] focused on ChatGPT's linguistic capabilities and the ethtical issues surrounding large language models. In [18], Zhuo et al. found that ChatGPT posed issues such as moral hazards, bias, reliability, robustness, and toxicity. They discussed several ethical challenges of advanced language models and offered suggestions for designing ethical models. The authors of [2] evaluated ChatGPT on various NLP tasks and found that it had limitations in understanding non-Latin script languages. They also found that it exhibited unreliable reasoning.

This collection of works shares with us the objective of understanding the performance of chatGPT and its underlying models (GPT-3 and GPT-4). That said. these works are so far apart from our research which is focused on the *coding* performance of ChatGPT.

## 3   DATA COLLECTION EXPERIMENTS

In this section we describe our data collection tools and environment, and provide details of the coding challenges presented to ChatGPT in our study.

## 3.1   Choice of tools used in our experiments

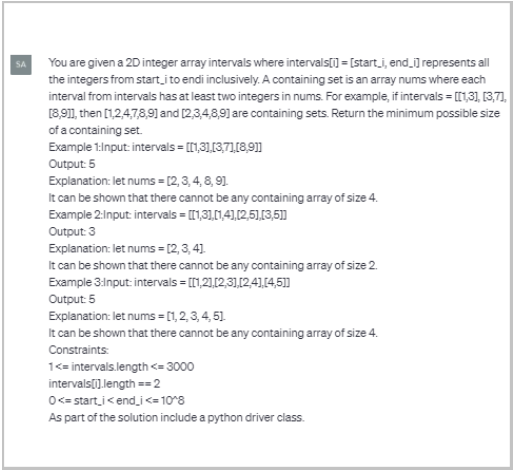
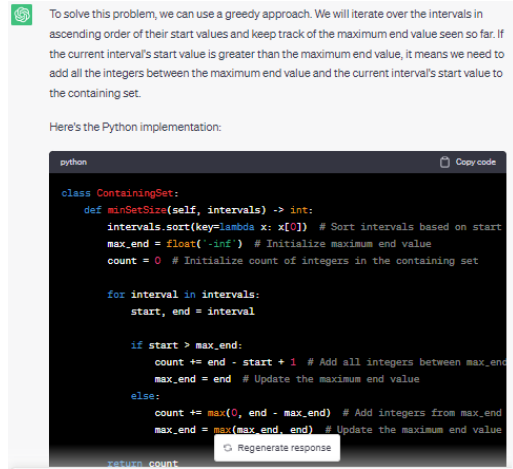

(a) A LeetCode question entered on the ChatGPT prompt

(b) Part of the code generated by ChatGPT in response to the question

Fig. 1.   Screenshot for one of the coding challenges posed to ChatGPT during our experiments

The two primary tools used during our ChaGPT evaluations were LeetCode [6] and Pylint [13]. LeetCode is an online platform with a wide range of coding challenges and interview preparation materials. It offers various problem difficulty levels across a wide range of topics and supports an extensive selection of programming languages. Most of the problems on LeetCode are inspired by, or directly drawn from interview questions at major tech companies such as Google, Amazon, Facebook, Microsoft, and so on. LeetCode contains a built-in compiler that not only assesses the code submitted by users but also compares it to other submissions. It evaluates code submissions based on an elaborate catalog of test cases. These exceptional features make LeetCode a solid platform for evaluating coding proficiency, and hence we use it in our experiments as both the source of coding challenges, and as a platform to evaluate the solutions presented to the challenges (See Figure 1 for an example of a LeetCode question submitted to ChatGPT, and the coding solution generated by ChatGPT during our experiments).

To evaluate the code quality of the programming solutions generated by ChatGPT, we used the Pylint [13] Python library. PyLint performs static analysis on Python code to identify adherence to coding standards and style guidelines, syntax, type, and logical errors, unused code, refactoring suggestions, etc. The comprehensiveness of code analysis provided by Pylint was the main reason for selecting this library in our experiments. It is noteworthy that the Pylint evaluations are undertaken in our experiments even when a ChatGPT code solution fails to solve the problem at hand. This is so because we are interested in examining the quality of ChatGPT's code regardless of whether a solution is wrong or not.

## 3.2   Data collection details

*3.2.1   The process.*   We copied each coding challenge from LeetCode and manually pasted it into the ChatGPT input screen. The manual approach allowed us to visually inspect each coding response produced by ChatGPT before pushing it further into our evaluation pipeline. Using the API would have made it difficult for us to flag cases where ChatGPT generated incomplete responses. For each coding solution generated by ChatGPT, we first input the code into LeetCode and recorded three pieces of information from the LeetCode output: (1) a binary value of whether the solution passes the question or not, (2) the proportion of human submissions that passed this question since it was published, (3) an error message if the code produces an error. We also input each coding solution into Pylint which analyzed code for quality issues and returned a problem type, ID for this problem type, and a textual description of the message (see Table 9 in Appendices).

*3.2.2 Experiment configurations.* Across all problems solved during our experiments, we had ChatGPT solve 2,792 coding challenges/questions, and hence generate 2,792 different python programs. About 52% (or 1,446) of these programs were part of what we refer to as the *complete coding challenges*. In these challenges, we entered each LeetCode question in its entirety into ChatGPT. In otherwards we entered the question, and supporting information provided by LeetCode, which includes constraints/conditions, and examples that clarify the question or nature of the expected solution. The 1,446 programs generated by ChatGPT during these complete coding challenges are further subdivided into two: 723 programs generated with GPT-3.5 as the underlying learning model, and 723 programs generated with GPT-4 as the underlying learning model. To enable comparison on an equal footing, the coding challenges submitted to GPT-3.5 were exactly the same as those submitted to GPT-4. This implies that the total number of *unique* coding challenges posed to ChatGPT in this portion of our experiment were 723.

The other 48% (or 1,346) of the challenges we posed to ChatGPT (and hence python programs generated by ChatGPT) were what we refer to as the *incomplete coding challenges*. Half (or 673) of the coding prompts were processed by GPT-3.5 and the other half by GPT-4. For the same reason previously discussed under the *complete coding challenges*, the questions posed to GPT-3.5 were also exactly the same as those posed to GPT-4. Hence the number of *unique* coding questions posed to ChatGPT in this portion of the experiment was 673.

In these incomplete coding challenges, the questions posed to ChatGPT had missing information. As mentioned in Section 1, the idea behind this experiment design was to investigate possibilities of ChatGPT memorizing parts of the training set. If ChatGPT can be found to generate correct source code for problems for which key information is missing, this could be strongly suggestive of possible memorization of training data that intersects with a portion of our questions and their answers. This could however also imply that ChatGPT is too smart that it intelligently fills in the missing information and successfully solves the incomplete question. We navigate the divide between these two possibilities by undertaking this code memorization experiment on both the train and test sets (see description of these two sets and their rationale in Section 3.2.4).

To collect data for this code memorization experiment, we input each question into ChatGPT, while leaving out the constraints and examples, two key pieces of information that not only clarify the problem, but also the nature of the expected solution. We use Figure 2 to illustrate our notion of complete and incomplete coding challenges. The figure shows a full LeetCode question that includes the problem text, a couple of examples and the constraints. Using this question in our incomplete coding challenges, we only entered the problem text into ChatGPT and left out the two examples, and the three constraints. In the complete coding challenges on the other hand, we entered everything shown in Figure 2 into ChatGPT.

*3.2.3 Selection of coding problems.* For reasons described in Section 1, the coding problems used in our experiments were on algorithms, data structures and strings. Table 2 shows how our problems were distributed across topics and sub-topics for both the complete and incomplete coding challenges. Observe that the number of problems on strings are far less than the number of problems from the other topics. We set these question proportions to approximately mirror the topic and sub-topic share of the LeetCode database. Table 3 shows the distribution of our questions across the difficulty levels specified on LeetCode. Observe that the questions having medium-level difficulty dominate our problem-set while the easy questions are fewest.

---

**Problem:** There are n projects numbered from 0 to n - 1. You are given an integer array milestones where each milestones[i] denotes the number of milestones the $i^{th}$ project has.

You can work on the projects following these two rules:
1. Every week, you will finish **exactly one** milestone of **one** project. You **must** work every week.
2. You **cannot** work on two milestones from the same project for two **consecutive** weeks.

Once all the milestones of all the projects are finished, or if the only milestones that you can work on will cause you to violate the above rules, you will stop working. Note that you may not be able to finish every project's milestones due to these constraints.

Return the **maximum number** of weeks you would be able to work on the projects without violating the rules mentioned above.

**Example 1:**
**Input:** milestones = [1,2,3]
**Output:** 6
**Explanation:** One possible scenario is:
- During the $1^{st}$ week, you will work on a milestone of project 0.
- During the $2^{nd}$ week, you will work on a milestone of project 2.
- During the $3^{rd}$ week, you will work on a milestone of project 1.
- During the $4^{th}$ week, you will work on a milestone of project 2.
- During the $5^{th}$ week, you will work on a milestone of project 1.
- During the $6^{th}$ week, you will work on a milestone of project 2.
The total number of weeks is 6.

**Example 2:**
**Input:** milestones = [5,2,1]
**Output:** 7
**Explanation:** One possible scenario is:
- During the $1^{st}$ week, you will work on a milestone of project 0.
- During the $2^{nd}$ week, you will work on a milestone of project 1.
- During the $3^{rd}$ week, you will work on a milestone of project 0.
- During the $4^{th}$ week, you will work on a milestone of project 1.
- During the $5^{th}$ week, you will work on a milestone of project 0.
- During the $6^{th}$ week, you will work on a milestone of project 2.
- During the $7^{th}$ week, you will work on a milestone of project 0.
The total number of weeks is 7.
Note that you cannot work on the last milestone of project 0 on $8^{th}$ week because it would violate the rules. Thus, one milestone in project 0 will remain unfinished.

**Constraints:**
n == milestones.length
1 <= n <= $10^5$
1 <= milestones[i] <= $10^9$

---

Fig. 2. Example of a full LeetCode question that illustrates the difference between our *complete* and *incomplete* coding challenges. In the incomplete coding challenges, the 2 examples and 3 constraints shown in the figure *were omitted* as only the problem text was input into ChatGPT. In the complete coding challenges, the problem text, 2 examples and 3 constraints were all input into ChatGPT.

This distribution of difficulty levels is again a design choice meant to ensure that the difficulty levels of our problems approximately reflect those of the LeetCode database.

Figure 3 provides an in-depth view of how the difficulty levels of our problems compare to those of the entire LeetCode database. The X-axis shows the acceptance rates of the questions, while the Y-axis shows the number of questions which had a given acceptance rate. The figure shows that the distribution of acceptance rates of our chosen questions closely aligns with that of the entire LeetCode database. Because LeetCode is in some ways now established as a reference standard for coding proficiency (e.g. for Big Tech interview prep), tailoring our problem distribution closely to patterns exhibited by the entire LeetCode database should enable the community to more rigorously contextualize our findings on ChatGPT's performance.

*3.2.4 Why we undertake evaluations on both the training and testing sets.* ChatGPT was trained on almost everything present on the internet before September 2021. When analyzing its performance therefore, its important to draw the line between information (e.g., questions and answers) that existed on the internet before that date and those that didn't. For most conventional machine learning applications pre-ChatGPT, the standard performance bechnmarking procedure is to only perform tests using data that did not feature in the training process. Due to the unique nature of ChatGPT's application scenarios however, one needs to perform both evaluations, albeit separately.

In particular, by virtue of ChatGPT being trained on almost the entirety of the internet, a significant proportion of the questions that will be posed by end-users to ChatGPT in real-life were possibly embedded in the training set along with their answers. For example, a student using ChatGPT as a tool to learn about coding will likely query it for a lot of code snippets that were posted on coding forums pre-ChatGPT in response to users' questions. Such questions and answers would hence have been directly embedded in the training set. ChatGPT performance assessments done using questions that existed on the internet along with their answers during the training phase might be more accurately reflective of how ChatGPT performs when faced with these kinds of queries.

On the other hand, assessments made using data (or queries) that never (directly) existed on the internet before the September 21 cut-off data might more accurately capture how ChatGPT will perform on previously unseen queries. Such queries would require ChatGPT to more intricately piece together various snippets of information in order to generate a correct solution. Both performance perspectives are however important to both the end-user and AI practitioner. This is because the effective performance exhibited by ChatGPT in the wild will likely sit in between these two extremes. In this paper we use the terms train set (or training set) and test set (or testing set) to respectively refer to queries that along with their solutions, directly existed on the internet before and after September 2021. Throughout, we will present and analyze results for each of these separately.

## 4 EXPERIMENTAL RESULTS

In this section, we present results for how ChatGPT performed across the full range of our coding experiments.

### 4.1 How often does ChatGPT produce a correct coding solution?

*4.1.1 Complete coding challenges.* For each of GPT-3, GPT-4 and human coders, Figure 4 shows the number of correct coding solutions expressed as a percentage of the total number of coding solutions generated. The correct solutions are tallied for all problems solved in our experiments without separating them into topics and sub-topics. The figure shows results for both train and test sets, and is specifically for only the complete coding challenges. For coding challenges drawn from the train set, the figure shows that GPT-4 clearly stands out, with GPT-3 performing about as well as humans. On the test set, humans outperform both models, while GPT-4 performs about twice as well as GPT-3.

Table 2. Percentage of LeetCode questions of different topics compared to the total no. and percentage of question no. of sub-topics compared to the topics they belong to in the dataset.

| Topic | No. of Questions (%) | | Sub-topic | No. of Questions (%) | |
|---|---|---|---|---|---|
| | Complete coding challenges | Incomplete coding challenges | | Complete coding challenges | Incomplete coding challenges |
| Algorithm | 422 (58.40%) | 407 (60.48%) | Dynamic | 132 (31.30%) | 124 (30.47%) |
| | | | Greedy | 136 (32.23%) | 129 (31.70%) |
| | | | Depth first search | 99 (23.46%) | 99 (24.32%) |
| | | | Divide and conquer | 33 (7.82%) | 33 (8.11%) |
| | | | Topological sort | 22 (5.21%) | 22 (5.41%) |
| Data Structure | 248 (34.30%) | 228 (33.88%) | Priority queue | 82 (33.06%) | 82 (35.96 %) |
| | | | Array | 49 (19.76%) | 45 (19.74 %) |
| | | | Hash table | 43 (16.94%) | 42 (18.42 %) |
| | | | Stack | 38 (15.73%) | 33 (14.47 %) |
| | | | Binary Search Tree | 36 (14.52%) | 26 (11.40 %) |
| Strings | 53 (7.30%) | 38 (5.65%) | | | |
| Total Questions | 723 | 673 | | | |

Table 3. Dataset overview based on difficulty levels of Leetcode questions.

| Difficulty Level | Overall question share(%) | | Topics | Number of questions | |
|---|---|---|---|---|---|
| | Complete coding challenges | Incomplete coding challenges | | Complete coding challenges | Incomplete coding challenges |
| **Easy** | 96 (13.30%) | 83 (12.31%) | Algorithm | 40 | 38 |
| | | | Data Structure | 42 | 39 |
| | | | Strings | 14 | 6 |
| **Medium** | 374 (51.7%) | 345 (51.18%) | Algorithm | 230 | 217 |
| | | | Data Structure | 126 | 116 |
| | | | Strings | 18 | 12 |
| **Hard** | 253 (35.00%) | 246 (36.49%) | Algorithm | 152 | 152 |
| | | | Data Structure | 80 | 74 |
| | | | Strings | 21 | 20 |

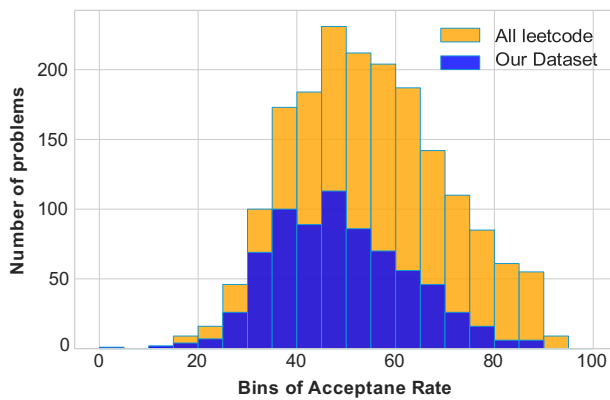

Fig. 3. A histogram showing how the acceptance rates of solutions to LeetCode questions used in our experiments compare to the acceptance rates across all problems hosted on the entire LeetCode platform. These acceptance rates are tracked by LeetCode when humans solve problems on the platform

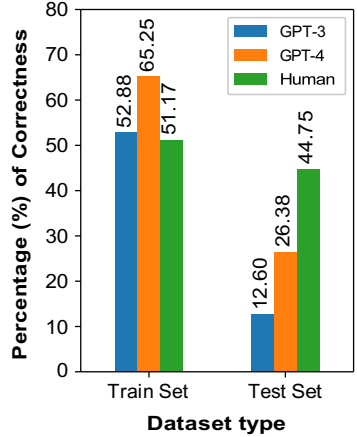

Fig. 4. Correctness of GPT-3, GPT-4 and Humans for Train and Test sets.

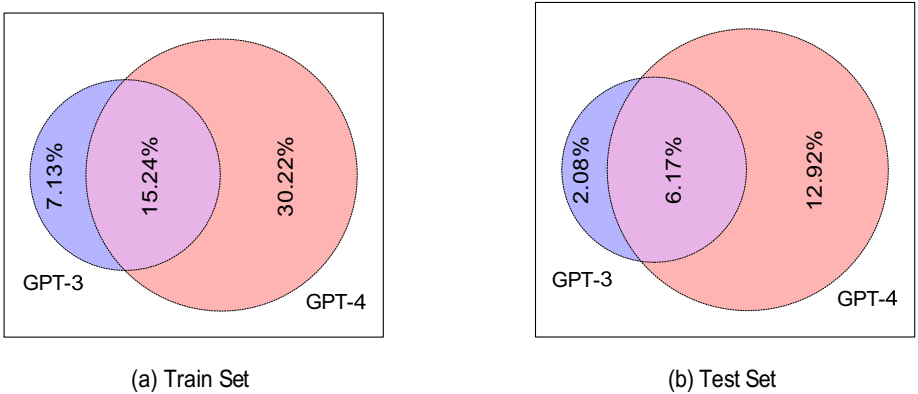

(a) Train Set                                          (b) Test Set

Fig. 5. Venn diagram showing exclusive and inclusive correctness of GPT-3 and GPT-4 for all the problems in the train and test datasets.

Comparing each model's performance between the train and test sets, the figure shows that GPT-3's effectiveness on the test set is almost 25% of that on the train set while GPT-4's effectiveness on the test set is about 50% of that on the train set. GPT-4 hence does a significantly better job at generalization than GPT-3. Overall, the takeaway from the figure is that GPT-4 is a significant improvement over GPT-3, and that for use-cases requiring coding solutions to algorithms and data structures problems different from those seen during training, humans continue to do much better than either ChatGPT model.

In light of GPT-4's architectural improvements over GPT-3 and the above discussed superior performance relative to GPT-3, it is instructive to pose the following question: did GPT-4 achieve its performance numbers by successfully solving *every* problem that GPT-3 solved, in addition to other problems that GPT-3 could not solve? We address this question in the venn diagrams shown in Figure 5. For both the train and test datasets, the figures show that there are some questions for which GPT-3 generates a correct solution yet GPT-4 generates an incorrect solution (i.e., 7.13% of the questions in the train set and 2.08% on of the questions on the test set). Hence while GPT-4 is clearly more effective than GPT-3, this does not mean that it beats GPT-3 on every single task. The trait points to a reality of learning-based systems where an enhanced model can generally deliver better performance overall but yet perform worse on certain problems. Note that we do not include human performance on Figure 5 because this information is not available to us at this granularity. In particular, LeetCode only publishes aggregate statistics for the percentage of correct solutions submitted by its user base for a given problem. Our venn diagrams on the other hand depict each of GPT-3 and GPT-4 as an individual user who either obtains a PASS or a FAIL for each question.

Table 4 and Table 5 dig deeper into the performance analysis by breaking it down to topic and sub-topic level. Table 4 shows that the patterns on the relative performance between models that we reported in Figure 4 persist even at topic-level since GPT-4 performs best on the train set while humans perform best on the test set *for each topic*. This general pattern still holds at the sub-topic level (see Table 5). Table 5 also suggests that some topics are markedly better done than others irrespective of the model in question. For example, on the training set, dynamic programming has our 3 entities (i.e., the two models and humans) at between 32% and 44% accuracy, while depth-first search has the 3 entities at between 57% and 87% accuracy. A similar phenomena can be seen across topics on the test set. As we discuss in the following paragraph, this trend is more a result of the difficulty of the questions that were solved within a topic as opposed to the identity of the topic itself.

Table 4. Correctness of GPT-3, GPT-4, and Humans for each problem topics in terms of Train and Test datasets.

|  | Topic | Correct Solutions % | |
| --- | --- | --- | --- |
|  |  | Train Set | Test Set |
| GPT-3 | Algorithms | 51.90 | 10.62 |
|  | Data Structures | 54.91 | 10.67 |
|  | Strings | 50.00 | 36.84 |
| GPT-4 | Algorithms | 66.79 | 23.12 |
|  | Data Structures | 63.00 | 29.34 |
|  | Strings | 64.70 | 42.10 |
| Human | Algorithms | 49.39 | 44.02 |
|  | Data Structures | 54.76 | 44.80 |
|  | Strings | 46.57 | 50.64 |

Table 6 shows how performance varied across difficulty levels of the coding challenges. Observe that, irrespective of the topic, the percentage of correct solutions drops as one navigates from the easy questions at the top of the table towards the hard questions lower down the table. The table confirms that the difficulty of the questions, and not the topic is the stronger determinant of what proportion of coding solutions are correct for both train and tests sets. Another interesting trait seen here is that humans hold better as the questions get harder. For example, on the test set, GPT-4 drops from a peak accuracy of 75% on the easy questions, to as low as 0% on the hard questions. Humans on the other see a more modest drop from 69.81% to 33.62% as questions on the test set get harder. On the training set, the best of the two models does significantly better than humans on the easy questions, yet the humans close the gap on the harder problems down the table.

4.1.2 *Incomplete coding challenges.* Figure 6 shows the performance registered with the incomplete coding challenges (recall Section 3.2.2). For context, we have included the performance numbers

Table 5. Performance of GPT-3, GPT-4 and humans for each sub-topic across the Train and Test datasets

| Sub-topic | Subjects | Correct Solutions % | |
|---|---|---|---|
| | | Train Set | Test Set |
| Dynamic Programming | GPT-3 | 32.43 | 6.90 |
| | GPT-4 | 44.59 | 15.52 |
| | Human | 43.73 | 39.41 |
| Greedy Algorithms | GPT-3 | 42.50 | 14.29 |
| | GPT-4 | 66.25 | 23.21 |
| | Human | 46.55 | 42.66 |
| Depth first search | GPT-3 | 77.94 | 16.13 |
| | GPT-4 | 86.76 | 38.71 |
| | Human | 56.89 | 53.50 |
| Divide and conquer | GPT-3 | 71.43 | 0 |
| | GPT-4 | 78.57 | 40.0 |
| | Human | 54.30 | 43.24 |
| Topological sort | GPT-3 | 41.67 | 0 |
| | GPT-4 | 66.67 | 10.0 |
| | Human | 49.35 | 49.40 |
| Priority queue | GPT-3 | 52.08 | 5.88 |
| | GPT-4 | 70.83 | 35.29 |
| | Human | 51.46 | 46.89 |
| Arrays | GPT-3 | 41.03 | 20.0 |
| | GPT-4 | 43.59 | 20.0 |
| | Human | 54.11 | 38.37 |
| Hash tables | GPT-3 | 59.38 | 9.09 |
| | GPT-4 | 56.25 | 30.0 |
| | Human | 50.23 | 49.16 |
| Stacks | GPT-3 | 57.14 | 20.0 |
| | GPT-4 | 71.43 | 27.27 |
| | Human | 56.72 | 44.96 |
| Binary Search Tree | GPT-3 | 73.08 | 10.0 |
| | GPT-4 | 76.92 | 20.0 |
| | Human | 65.30 | 39.67 |

Table 6. Performance of GPT-3, GPT-4 and humans across problems having varying difficulty levels

| Difficulty Level | Topics | Correct Solutions % | | | | | |
|---|---|---|---|---|---|---|---|
| | | Train Set | | | Test Set | | |
| | | GPT-3 | GPT-4 | Humans | GPT-3 | GPT-4 | Humans |
| Easy | Algorithms | 96.42 | 96.42 | 58.88 | 33.34 | 58.34 | 55.95 |
| | Data Structures | 72.72 | 81.81 | 68 | 33.34 | 66.67 | 55.64 |
| | Strings | 100 | 83.33 | 52.4 | 62.5 | 75 | 69.81 |
| Medium | Algorithms | 58.15 | 75.17 | 51.17 | 14.6 | 30.34 | 44.93 |
| | Data Structures | 64.04 | 64.04 | 54.5 | 8.1 | 35.13 | 45.7 |
| | Strings | 54.54 | 72.72 | 46.89 | 28.57 | 28.57 | 38.47 |
| Hard | Algorithms | 29.03 | 45.16 | 43.83 | 0 | 5.08 | 40.22 |
| | Data Structures | 27.45 | 49.01 | 46.64 | 6.45 | 10.34 | 40.31 |
| | Strings | 29.41 | 52.94 | 44.31 | 0 | 0 | 33.62 |

(blue and green bars) from the earlier presented results on the complete coding challenges. On the training set (Figure 6a), the figure shows that irrespective of whether the coding questions are complete or not, GPT-3 attains about the same performance. The same trend is seen with GPT-4. Also noteworthy is that humans, despite having the benefit of complete information on all the coding challenges, performed worse than both GPT-3 and GPT-4 models which operated with incomplete information, There are two possible explanations for the patterns seen here. First, it could be that the GPT-3 and GPT-4 models memorized these problems and their answers from the training dataset, enabling them to perform the same way even when significant information was cut our from the questions. It could however also be that the two GPT models have built a good amount of domain knowledge from the vast expanse of training data, enabling them to fill in the gaps where coding questions are missing the kinds of information specified in Section 3.2.2.

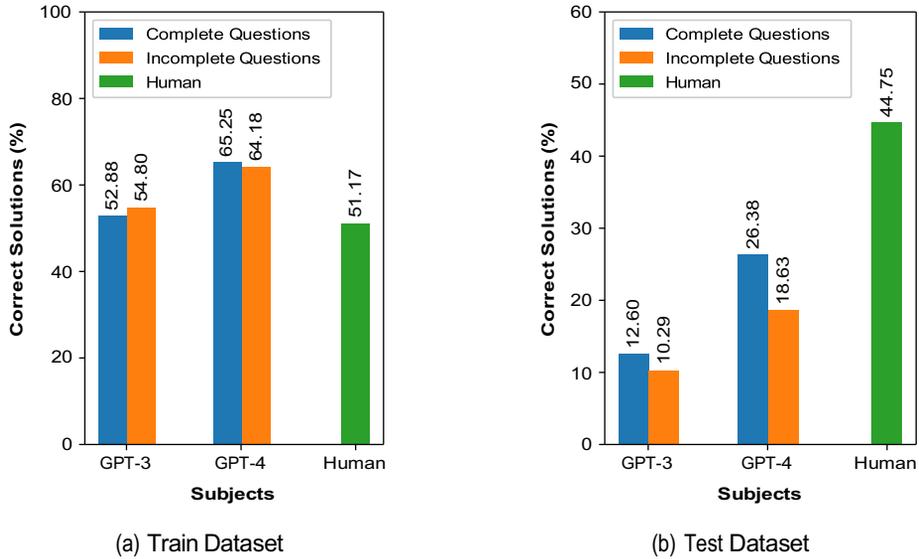

(a) Train Dataset  (b) Test Dataset

Fig. 6. Performance of GPT models on the incomplete coding challenges

Figure 6b provides some clarity on the likelihood of each of the above two possible explanations. The figure shows how the models performed when the incomplete coding challenges were drawn from the test set. As has been seen with all test sets up to this point, the percentage of correct solutions for both models is markedly reduced relative to when the training set was used. That recurring observation aside, it is noteworthy that GPT-3 and GPT-4 still get between 10% and 19% of the incomplete coding questions correct. This being a test set, the performance seen here cannot be attributed to memorization given that the test questions/answers were not included in the data used to train the model. A more plausible explanation is that the the 10-19% correct solutions were due to the models filling in the gaps for the missing constraints, examples and diagrammatic illustrations. The sum-total of behavior registered in Figures 6a and 6b could hence be attributed to a bit of both memorization and robustness of the two GPT models.

### 4.2 Examining the cases when ChatGPT failed to produce a correct solution

Up to this point, we have focused on the percentage of correct coding solutions obtained across a wide range of scenarios. It is however insightful to also dig into the cases when LeetCode found the solutions to be incorrect. In particular, we are interested in those cases where ChatGPT produced incorrect solutions whose code executed successfully (as opposed to (incorrect) solutions which could not even run). It is interesting to get some measure of whether these wrong solutions whose code executed successfully were only marginally wrong or so wrong.

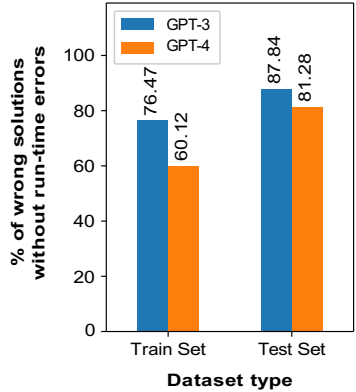

Fig. 7. Percentage of problems produced errors, which indicates the percentage of errors out of all incorrect answers (errors and wrong answers).

Figure 7 provides information on how frequently these cases occurred. The y-axis shows the number of cases when ChatGPT produced an incorrect solution whose code executed successfully (i.e., code that did not produce a run-time error) expressed as a percentage of all the cases when it failed to produce a correct solution (i.e., the total of the cases when it produced an incorrect solution whose code executed successfully and those when the code returned a run-time error). On the training set, the figure shows that ChatGPT produced between 60% and 77% of such cases. On the test set, these cases happen 80% of the time or more. Overall, the graph shows that in the majority of cases when ChatGPT failed to produce a correct solution, the code still executed successfully. The figure also shows that on both the train and test set, GPT-4 was less likely to generate wrong solutions that had no run-time errors (i.e., GPT-4's wrong solutions were more likely to generate errors than GPT-3's wrong solutions). This trait is somewhat surprising since all our results show GPT-4 to clearly be the superior model when seen from the perspective of the proportion of correct solutions it generates. An examination of the nature of errors in the cases where ChatGPT produced run-time errors will be done in Section 4.3.

Here, we dig deeper into the cases where ChatGPT's code executed successfully despite the solutions being wrong. Table 7 sheds light on the following question: *how wrong were ChatGPT's wrong solutions?* The relevance of this question lies in the fact that, if the majority of wrong solutions turn out to only be *marginally wrong*, then users in practice might perhaps simply have to make minor tweaks to these wrong solutions in order to solve their coding challenges. To answer the question, we leverage the test cases which have been meticulously put together in LeetCode for each coding challenge. When a solution is flagged as wrong, LeetCode still returns the number of test cases passed. We use the fraction of test cases passed as some measure of how wrong a given solution is — i.e., we classify a solution with a smaller proportion of test cases passed as *more wrong* than one which has a larger proportion of test cases passed.

Table 7 shows the percentage of test cases that were passed by the wrong solutions generated by ChatGPT for various experiment configurations. The percentages of test cases (shown in the first column) have been organized in bins of width 10 from 0-10 to 90-100. To illustrate how to read this table, we use the first value (42.57) in the second column (that is, under Algorithms -> Train set -> GPT-3). This value means that 42.57% of the wrong answers generated by the GPT-3 model passed between 0 to 10% of the test cases when coding challenges for algorithms problems were drawn from the train set. Looking at the first two rows of the entire table therefore, one can conclude that the vast majority of the wrong solutions generated by ChatGPT passed only 0-20% of the test cases. At the other extreme (last row), one can see that a very small proportion of the wrong answers passed 90% to just under 100% of the test cases. Overall, the table reveals that for the kinds of coding problems solved in this paper, whenever ChatGPT generated a wrong solution, the solution was, based on test cases passed, very likely to be far detached from the correct solution. The table shows that for the most part this pattern still holds regardless of the model (GPT-3 vs GPT-4), problem topic (algorithms vs data structures vs strings), and the source of the problems (i.e., train set vs test set).

Table 7. Percentage of test cases passed when ChatGPT generated wrong solutions

| % of test cases passed | % of wrong solutions that passed the given % of test cases | | | | | | | | | | | |
|---|---|---|---|---|---|---|---|---|---|---|---|---|
| | Algorithms | | | | Data structures | | | | Strings | | | |
| | Train Set | | Test Set | | Train Set | | Test Set | | Train Set | | Test Set | |
| | GPT-3 | GPT-4 | GPT-3 | GPT-4 | GPT-3 | GPT-4 | GPT-3 | GPT-4 | GPT-3 | GPT-4 | GPT-3 | GPT-4 |
| (0, 10] | 42.57 | 45.45 | 54.40 | 49.51 | 35.19 | 37.84 | 43.33 | 37.78 | 50.00 | 66.67 | 44.44 | 25.00 |
| (10, 20] | 16.83 | 18.18 | 16.00 | 15.53 | 18.52 | 18.92 | 16.67 | 22.22 | 0.00 | 16.67 | 0.00 | 25.00 |
| (20, 30] | 8.91 | 7.27 | 7.20 | 6.80 | 1.85 | 16.22 | 5.00 | 2.22 | 0.00 | 0.00 | 0.00 | 0.00 |
| (30, 40] | 2.97 | 1.82 | 3.20 | 3.88 | 1.85 | 5.41 | 8.33 | 8.89 | 0.00 | 0.00 | 22.22 | 0.00 |
| (40, 50] | 5.94 | 3.64 | 1.60 | 6.80 | 3.70 | 8.11 | 5.00 | 8.89 | 8.33 | 0.00 | 11.11 | 0.00 |
| (50, 60] | 0.99 | 1.82 | 4.00 | 2.91 | 7.41 | 0.00 | 1.67 | 6.67 | 0.00 | 0.00 | 11.11 | 0.00 |
| (60, 70] | 3.96 | 3.64 | 4.80 | 2.91 | 3.70 | 5.41 | 3.33 | 2.22 | 0.00 | 0.00 | 0.00 | 50.00 |
| (70, 80] | 6.93 | 3.64 | 4.00 | 6.80 | 11.11 | 5.41 | 5.00 | 6.67 | 8.33 | 0.00 | 11.11 | 0.00 |
| (80, 90] | 3.96 | 1.82 | 3.20 | 1.94 | 9.26 | 0.00 | 5.00 | 2.22 | 16.67 | 0.00 | 0.00 | 0.00 |
| (90, 100) | 6.93 | 12.73 | 1.60 | 2.91 | 7.41 | 2.70 | 6.67 | 2.22 | 16.67 | 16.67 | 0.00 | 0.00 |

## 4.3 Evaluation of ChatGPT's code quality

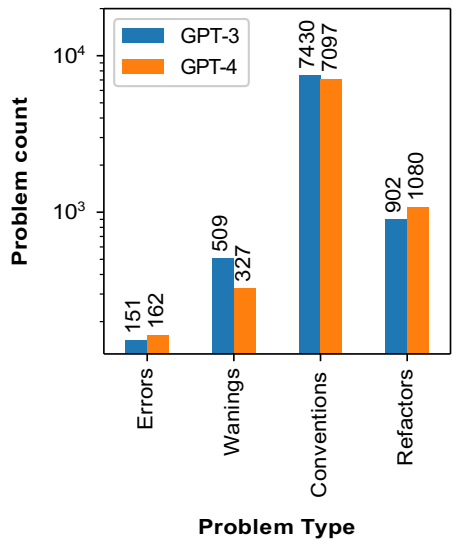

Fig. 8. Code quality issues seen in ChatGPT code

In this section, we present our findings on the code quality evaluations. We undertook these using Pylint [13], and performed evaluations on all coding solutions generated by ChatGPT regardless of whether they were wrong or correct. We also ran these evaluations for cases when ChatGPT's code generated run-time errors. In general, the notion of code quality remains relevant irrespective of whether a piece of code ultimately solves the problem at hand or not. The section particularly offers insights into the nature of errors, warnings, refactors, and convention violations seen in ChatGPT's code. Figure 8 shows the total count seen in our experiments for each of them.

We obtained a total of just over 300 errors, much lower in number than the other issues. On the other extreme, convention-related issues occurred over 15,000 times. In the following paragraphs, we break down each of these into their sub-types by percentage. The absolute numbers shown in Figure 8 should help provide context to the percentage numbers reported in the rest of the narrative.

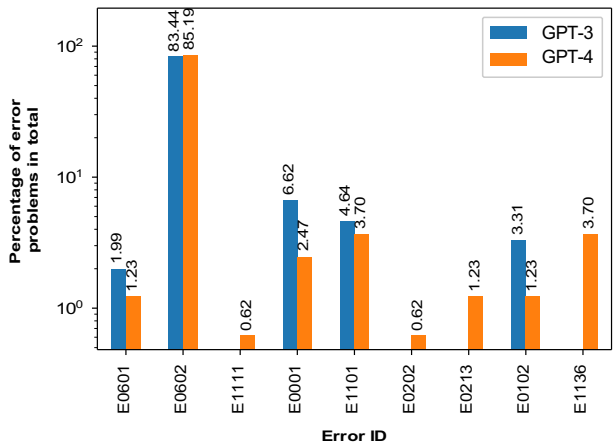

Fig. 9. Percentage share of each of the error types seen in ChatGPT code

Figure 9 shows the percentage share of each of the error types seen in our experiments. The figure shows that the error, E0602, occurred over 80% of the time for both GPT-3 and GPT-4; in fact, it occurred an entire order of magnitude more frequently than the other errors. It is for this reason that we used a logarithmic scale on the y-axis. According to the Pylint documentation [13], this error message is defined as *undefined-variable*, and is generated when a variable that was not defined is accessed. To the ChatGPT end-user, this finding suggests that whenever ChatGPT-generated code returns run-time errors, a careful review and fixing of variable definitions might be all that one needs to solve many of the problems. Each of the other error types occur so infrequently and we dont discuss them here. However, in the appendix section (see Appendix A, Page 22), we provide a definition for the meaning of each of the errors seen in our study.

Figure 10 shows all the warnings which we observed in our experiment (13 in total). Again, we see a similar phenomena: one warning type occurs an entire order of magnitude more frequently than each of the others. In this case that warning type is, W0621, which Pylint defines as, *redefined-outer-name*. The warning is raised when one redefines a name from an outer scope. For example, if a variable was defined in a scope external to some function, this warning is raised when the same variable is defined inside the function. Looking at this dominant warning in light of the dominant error described in the previous paragraph, one can conclude that defining variables in a consistent manner is in general a standout problem exhibited by ChatGPT code. For meanings of the less frequent warnings, the reader is referred to Appendix A, Page 22.

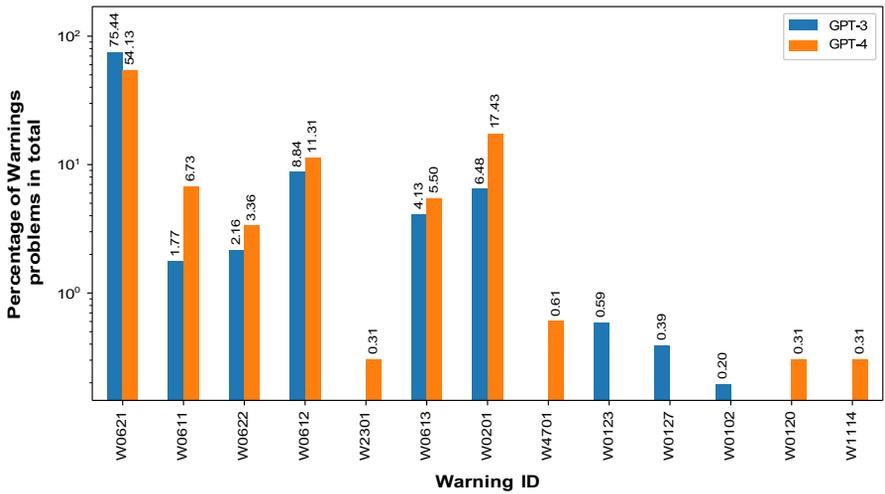

Fig. 10. Percentage share of each of the warning types seen in ChatGPT code

Finally, we found a total of 20 different refactor and 18 convention messages in our Pylint reports. These messages and their percentage of occurrence for both GPT-3 and GPT-4 can be seen in Table 8. Again we had a single refactor message and a single convention message occurring at a disproportionately high frequency relative to the others. The refactor message in question is the one with the message id *R0903*. According to the Pylint documentation, the message id refers to the message: *too-few-public-methods*. This is indicated when a class has a small number of public methods. On the other hand the dominant convention message is one with the message id *C0103*. It occurs when the given name does not adhere to the naming conventions specific to its type, whether it be a constant, variable, class, or otherwise. To the ChatGPT end-user who is leveraging it to generate source code, the overarching observation from this section is that one can expect just a couple of quality issues to occur the vast majority of times.

## 5 CONCLUSION

In this paper, we have conducted what is to date the most comprehensive evaluation of ChatGPT's programming proficiency. Our emphasis has been on algorithms and data structures, two topics at the heart of Computer Science. Using a total of 2,792 coding prompts, we have presented results from several performance perspectives. First, we have evaluated the proportion of correct solutions provided by ChatGPT across a wide range of sub-topics within the core topics. We have that humans passed more questions than ChatGPT on problems not seen in the training set, and fewer questions than ChatGPT on problems seen in the training set. We also found that despite the newer ChatGPT model (GPT-4) performing better than the older one (GPT-3), this is not necessarily the case on every problem. In particular, GPT-3 was able to outperform GPT-4 on several problems. This trait underlines one challenge that applies to learning algorithms in general, where enhancements to a learning model might make the model perform worse on certain problems, despite the model doing better overall. This observation suggests that as LLMs continue to advance, older versions of them might continue to be useful for a long while.

Table 8. Percentage share of each of the refactor and convention issues seen in ChatGPT code

| Percentage of lint-problems occurring in terms of all lint-problems | | | | | |
|---|---|---|---|---|---|
| Refactor | | | Convention | | |
| Message ID | GPT-3 (%) | GPT-4 (%) | Message ID | GPT-3(%) | GPT-4(%) |
| R0903 | 87.47 | 88.89 | C0103 | 46.97 | 47.53 |
| R1705 | 4.1 | 4.63 | C0116 | 14.41 | 15.63 |
| R0914 | 2.11 | 1.02 | C0115 | 12.36 | 14.79 |
| R1714 | 0.67 | 0.74 | C0114 | 9.6 | 10.13 |
| R0913 | 1.44 | 0.65 | C0303 | 14.24 | 7.69 |
| R1710 | 0.22 | 0.65 | C0301 | 1.41 | 2.49 |
| R1702 | 0.11 | 0.65 | C0200 | 0.74 | 0.63 |
| R1721 | 0.55 | 0.56 | C0321 | 0 | 0.45 |
| R1728 | 0.33 | 0.56 | C0305 | 0.15 | 0.39 |
| R0912 | 0.55 | 0.37 | C0121 | 0 | 0.08 |
| R1735 | 0.11 | 0.37 | C0415 | 0 | 0.06 |
| R1704 | 0.11 | 0.28 | C0325 | 0 | 0.06 |
| R0911 | 0 | 0.19 | C0206 | 0.07 | 0.03 |
| R1716 | 0.89 | 0.09 | C1803 | 0 | 0.01 |
| R0916 | 0.33 | 0.09 | C0413 | 0 | 0.01 |
| R1724 | 0.22 | 0.09 | C0304 | 0.03 | 0 |
| R1723 | 0.11 | 0.09 | C0209 | 0.01 | 0 |
| R1719 | 0.11 | 0.09 | C0201 | 0.01 | 0 |
| R1731 | 0.44 | 0 | | | |
| R0205 | 0.11 | 0 | | | |

Whenever ChatGPT provided a wrong solution, we proceeded and dug deeper to asses how many test cases the solution passed. This analysis enabled us to get some rough measure of how wrong each solution is, a kind of insight that end-users having to fix wrong ChatGPT code should find so useful. Our findings here indicate that the vast majority of ChatGPT's wrong solutions pass only a very small proportion of test cases.

Regardless of whether ChatGPT successfully solved a problem or not, we assessed the quality of code it generated. For this evaluation we looked into types of errors, warnings, conventions and refactor issues. We found that only a couple of error types, warnings, conventions and refactor issues show up the vast majority of times. For example, the single most frequent error type showed up an order of magnitude more frequently than each of the other error types. This same pattern, seen also with the other code quality attributes, suggests that end-users fixing ChatGPT code might have to focus on seeking out a small selection of issue types. In the training memorization experiment, we found evidence that suggests that some of the solutions generated by ChatGPT benefited from a fair amount of memorization of the training set. The same experiment however also provided results that suggested that part of what seemed like memorization might in fact be reflective of ChatGPT being so robust that it is able to fill in missing gaps in incomplete questions, thereby solving them successfully. This is an interesting question that surely call for more research.

Overall, while our results point out some issues where the latest iterations of LLMs (ChatGPT in particular) might still need to improve, the paper showcases the impressive performance of these models.

# A APPENDIX

Table 9. Definitions for all code quality issues seen in the ChatGPT code generated in our experiments

| Problem Type | Message ID | Message |
|---|---|---|
| Errors | E0601 | Using variable before assignment |
| | E0602 | Undefined variable |
| | E0001 | syntax error raised for a module; message varies |
| | E1101 | %s %r has no %r member |
| | E0213 | Method should have "self" as first argument |
| | E0102 | %s already defined line %s |
| | E1136 | Value '%s' is unsub scriptable |
| | E1111 | Assignment from no return |
| | E0202 | Method hidden |
| Warnings | W2301 | Unnecessary ellipsis |
| | W0621 | Redefining name %r from outer scope (line %s) |
| | W0611 | Unused import |
| | W0622 | Redefining built-in %r |
| | W0612 | Unused variable |
| | W0613 | Unused argument |
| | W0201 | Attribute defined outside __init__ |
| | W4701 | Modified iterating list |
| | W0123 | Eval used |
| | W0127 | Self assigning variable |
| | W0102 | Dangerous default value |
| | W0120 | Useless else on loop |
| | W1114 | Arguments out of order |
| Refactors | R0903 | Too few public methods |
| | R1705 | Unnecessary "%s" after "return" |
| | R1714 | consider using in |
| | R1710 | inconsistent return statements |
| | R0914 | too many locals |
| | R1721 | Unnecessary comprehension |
| | R1704 | Too many nested blocks |
| | R1702 | Too many nested blocks |
| | R1723 | No else break |
| | R1719 | Simplifiable if expression |
| | R1724 | No else continue |
| | R0912 | Too many branches |
| | R0911 | Too many return statements |
| | R1735 | Use dict literal |
| | R1728 | Consider using generator |
| | R0913 | Too many arguments |
| | R0205 | Useless object inheritance |
| | R0916 | Too many boolean expressions |
| | R1716 | Chained comparison |
| | R1731 | Consider using max builtin |
| Conventions | C0114 | Missing module docstring |
| | C0103 | %s name "%s" doesn't conform to %s |
| | C0115 | Missing class docstring |
| | C0116 | Missing function or method docstring |
| | C0303 | Trailing whitespace |
| | C0301 | Line too long |
| | C0200 | Consider using enumerate |
| | C0321 | Multiple statements |
| | C0305 | Trailing newlines |
| | C0304 | Missing final newline |
| | C0415 | Import outside toplevel |
| | C0206 | Consider using dict-items |
| | C0121 | Singleton comparison |
| | C1803 | Use implicit booleaness not comparison |
| | C0325 | Superfluous parens |
| | C0209 | Consider using f-string |
| | C0413 | Wrong import position |
| | C0201 | Consider iterating dictionary |